\documentclass[floatfix,twocolumn,aps,prl,amsmath,amssymb]{revtex4}
\usepackage[latin1]{inputenc}
\usepackage[dvips]{graphicx}
\usepackage{dcolumn}
\usepackage{bm}
\usepackage{color}
\usepackage{url}
\usepackage[colorlinks=true ,citecolor=blue]{hyperref}
\usepackage{amsmath,amssymb}

\begin{document}

\title{Magnetic Field Effects on the Transport Properties of High-Tc Cuprates  }
\author{E. C. Marino$^{1}$}\thanks{marino@if.ufrj.br}
\author{R. Arouca$^{1,2}$}\thanks{arouca@pos.if.ufrj.br}
\affiliation{$^1$Instituto de F\' isica, Universidade Federal do Rio de Janeiro, C.P. 68528, Rio de Janeiro, RJ, 21941-972, Brazil.}%
\affiliation{$^2$Institute for Theoretical Physics, Center for Extreme Matter and Emergent Phenomena, Utrecht University, Princetonplein 5, 3584 CC Utrecht, The Netherlands.}%Lines break automatically or can be forced with \\

\date{\today}% It is always \today, today,

\begin{abstract}
Starting from a recently proposed comprehensive theory for the high-Tc superconductivity in cuprates, we derive a general analytic expression for the planar resistivity, in the presence of an applied external magnetic field $\textbf{H}$ and explore its consequences in the different phases of these materials. As an  initial probe of our result, we show it compares very well with experimental data for the resistivity of LSCO at different values of the applied field. We also apply our result to Bi2201 and show that the magnetoresistivity in the strange metal phase of this material, exhibits the $H^2$ to $H$ crossover, as we move from the weak to the strong field regime. Yet, despite of that, the magnetoresistivity does not present a quadrature scaling. Remarkably, the resistivity H-field derivative does scale as a function of $\frac{H}{T}$, in complete agreement with recent magneto-transport measurements made in the strange metal phase of cuprates \cite{Hussey2020}. We, finally, address the issue of the $T$-power-law dependence of the resistivity of overdoped cuprates and compare our results with experimental data for Tl2201. We show that this provides a simple method to determine whether the quantum critical point associated to the pseudogap temperature $T^*(x)$ belongs to the SC dome or not.
\end{abstract}

\maketitle

{\bf 1) Introduction}\\
\bigskip

Any complete theory for superconductivity in the high-Tc cuprates must be capable to describe, besides the superconductivity mechanism itself, the properties of their normal phases.
The comprehension of such phases of the cuprates, actually, seems to be as challenging as that of the superconducting phase itself.

An interesting issue, in connection to this, is the range of different functional dependences on the temperature, which are exhibited by the resistivity as we cross the $T_c(x)$ superconducting (SC) dome. These are usually of the form
$\rho(T)\propto T^{1+\delta}$, where apparently $\delta\in [0,1]$. The precise value of $\delta $, however, is strongly dependent on the specific region of the SC dome where we cross the SC transition and, consequently, the previously vanishing resistivity acquires a temperature dependence.

The situation becomes even richer, when we apply an external magnetic field and consider the resistivity dependence on it. Then, a wide range of effects can be observed, including the destruction of the superconducting phase. 

A particularly interesting non-superconducting phase of the cuprates is the so-called Strange Metal (SM) phase \cite{Hussey2020, taillefer2010scattering, hu2017universal, ando2000carrier, ando2004electronic, gurvitch87, keimer2015quantum, varma89, varma99, faulkner2010strange, zaanen14, Sachdev18, zaanen2004superconductivity, legros2019universal, zaanen19planckian, sachdev97, sachdev_2011, phillips05, hfreire}, where the resistivity grows linearly with the temperature, with a slope that decreases with doping, proportionally to the pseudogap temperature $T^*(x)$ \cite{Res}. 
Recent studies reveal, however, that specially in the case of overdoped (OD) cuprates \cite{HusseyTl}, depending on the doping amount, we not always move directly from the SC phase to a linear dependent resistivity. In many cases, for some compounds, we rather observe a super-linear dependence on $T$ before we reach the linear regime \cite{Hussey2020}.

Interesting experimental studies have also addressed the issue of the effect of an external magnetic field on the transport properties of OD cuprates \cite{Hussey2020}.
Such studies reveal, for instance, the existence of a crossover in the magnetic field dependence of the magnetoresistivity (MR) in the SM phase, ranging from a  quadratic behavior, at weak fields, to a linear one, in the strong field regime \cite{Hussey2020}. Such a behavior is analogous to the one observed in quantum critical phases of electron doped cuprates \cite{eldopcup} and pnictide superconductors  \cite{pnic,gg}.

In such systems, the crossover was ascribed to a quadrature scaling behavior, in which the planar MR behaves according to the empirical expression
\begin{equation}
    \rho(T,H)-\rho(0,0)=\sqrt{ \left (\alpha k_BT\right )^2 + \left (\gamma \mu_B\mu_0 H\right )^2}
    \label{quad}
\end{equation}
where $\alpha$ and $\gamma$ are constant fitting parameters.

A benchmark of the quadrature behavior is that the quantity $\Delta\rho/T=\left (\rho(T,H)-\rho(0,0)\right )/T$ becomes a function of the ratio $H/T$, namely
\begin{equation}
    \left (\rho(T,H)-\rho(0,0)\right )/T \propto\sqrt{ 1 + \left (\lambda \frac{ \mu_B\mu_0 H}{k_BT} \right )^2}.
    \label{quad1}
\end{equation}

The study carried on in \cite{Hussey2020} on the cuprates Bi2201 and Tl2201 shows that in spite of exhibiting the $H^2$ to $H$ crossover in the MR field dependence, the MR data for cuprates in the SM phase do not scale as the quadrature would do, namely, as in (\ref{quad1}).

Interestingly and remarkably, however, it was shown in \cite{Hussey2020} that the MR data for the resistivity field derivative, $(\partial\rho(T,H)/\partial H)$, do scale  as in (\ref{quad1}), namely,
\begin{equation}
 \frac{\partial\rho(T,H)}{\partial H} =f\left( \frac{H}{T}\right ).
    \label{quad2}
\end{equation}

In two recent publications \cite{marino2020superconducting,Res} we developed a comprehensive theory for the high-Tc cuprates, whose most distinguishable feature, perhaps, is to be testable. Indeed, our theory allows for the theoretical determination of several physical quantities, which can be directly compared with the experiments. Among these, we have obtained analytical expressions for the superconducting (SC) and pseudogap (PG) transition temperatures $T_c$ and $T^*$ as a function of quantities such as the stoichiometric doping parameter, number of planes, pressure and external magnetic field \cite{marino2020superconducting,Res}. We have also obtained a general expression for the resistivity as a function of the temperature in the different non-superconducting phases of the high-Tc cuprates \cite{Res}.These results are in excellent agreement with the experiments for a wide range of cuprate systems with one, two and three planes per unit cell.

In this work, we directly derive from the aforementioned theory, a general expression for the planar resistivity as a function of an applied external magnetic field $H$.

We firstly apply this result in order to describe the resistivity in LSCO and specially to determine how it is modified when the system is under the action of an external magnetic field.

We, then, consider our
expression for the resistivity in the SM phase and show that, interestingly, our expression completely agrees with the experimental results found in \cite{Hussey2020} for Bi2201. In particular, it exhibits the $H^2$ to $H$ crossover, in spite of the fact that it does not present the quadrature scaling behavior. Yet, it satisfies the field derivative scaling (\ref{quad2}).

Finally we address the issue of the super-linearity of the resistivity of OD cuprates, right above the the SC transition and offer a simple explanation, which is illustrated by comparison with experimental data for Tl2201.\\
\bigskip

{\bf 2) The Resistivity }\\
\bigskip

{\bf 2.1) General Expression}\\
\bigskip

The resistivity can be obtained as the inverse conductivity matrix, which is given by the Kubo formula
\begin{eqnarray}
\sigma^{ij}_{\text{DC}}=\lim\limits_{\omega\rightarrow 0}\frac{i}{ \omega}\left[1- e^{-\beta\hbar\omega} \right ]\lim\limits_{\mathbf{k}\rightarrow \mathbf{0}} \Pi^{ij}\left(\omega + i\epsilon, \mathbf{k}\right),
\label{10z}
\end{eqnarray}
where $\Pi^{ij}$ is the retarded, connected current-current correlation function:
\begin{eqnarray}
\Pi^{ij}=\langle j^{i}j^{j}\rangle_{\text{C}}.         
\end{eqnarray}
This is given by the second functional derivative of the grand-canonical potential in the presence of an applied electromagnetic vector potential $ \textbf{A}\left(\omega, \mathbf{k}\right)$, namely, 
\begin{eqnarray}
\langle j^i j^j\rangle_C \left(\omega, \mathbf{k}\right) = \frac{\delta^2  \Omega [ \textbf{A} ] }{\delta \textbf{A}^i\left(\omega, \mathbf{k}\right) \delta \textbf{A}^j\left(\omega, \mathbf{k}\right)},
\label{eq_j}
\end{eqnarray}

$\Omega [ \textbf{A} ]$ relates to the grand-partition functional $Z[\textbf{A}]$ as
\begin{eqnarray}
\Omega [ \textbf{A} ]=-\frac{1}{\beta} \ln Z[\textbf{A}] ,
 \label{2}
\end{eqnarray}
which is given by
\begin{eqnarray}
Z[\textbf{A}]= {\rm Tr}_{Total} e^{-\beta \left[ H[\textbf{A}]-\mu\mathcal{N}\right ]}.
 \label{3}
\end{eqnarray}
	In the expression above, $ H[\textbf{A}]$, is our proposed Hamiltonian for the cuprates\cite{marino2020superconducting,Res}, in the presence of an external field 
	 \begin{equation}
    \textbf{A}=\frac{1}{2} \textbf{r}\times \textbf{B}, 
\end{equation}
which corresponds to a constant external magnetic field $\textbf{B}= \mu_0 \textbf{H}$.

The trace above can be evaluated with the help of the eigenvalues of $ H[\textbf{A}]-\mu\mathcal{N}$, which are given by \cite{marino2020superconducting,Res}

 \begin{eqnarray}
 \mathcal{E}_{l}[\textbf A]=\sqrt{\Delta^2+\Big (\sqrt{v^2 (\hbar\textbf k +e \textbf A)^2 + M^2} + l\mu \Big )^2},
\label{7}
\end{eqnarray}
where $l=\pm 1$. The expression above is given
 in terms of the external field 
 and the ground-state expectation values: of the Cooper pair operator, $\Delta$, of the exciton operator, $M$ and of the chemical potential, $\mu$. The field dependence is conveniently expressed through the replacement
\begin{eqnarray}
  M^2\longrightarrow M^2 +2 e v\ \hbar \textbf{k}\cdot\textbf{A} + e^2v^2 \textbf{A}^2,
  \\ \nonumber
   M^2\longrightarrow M^2 -2 e v\  \langle \textbf{L}\rangle\cdot\textbf{H} + \frac{1}{4}e^2v^2 \langle r^2 \rangle \textbf{H}^2
\label{7aaa}
\end{eqnarray}
 in the presence of an applied field, where we replaced $r^2$ and $\textbf{L}$ for their average values. Since the ground state, either $|p_x\rangle$ or $|p_y\rangle$ is a linear combination of $|l,m\rangle = |1,\pm 1\rangle$, it follows that $\langle L_z\rangle=0$ and the second term in (\ref{7aaa}) does not contribute, thus confirming the observation made in \cite{Hussey2020} that there is no contribution of the orbital coupling with the external field. This also leads to results that are independent of the specific direction of the applied external magnetic field, which is in agreement with the experimental observations reported in \cite{Hussey2020}.

The grand-partition functional $Z[\textbf{A}]$ follows from  Eq.~\eqref{3} and Eq.~\eqref{7}, and after functional integration over the fermionic (holes),
degrees of freedom, namely \cite{marino2020superconducting,Res}
\begin{widetext}
\begin{eqnarray}
Z[\textbf{A}]&=&\exp\left\{-\beta\left\{  \frac{\left|\Delta\right|^2}{g_S}+\frac{\left|M\right|^2}{g_P}+N\mu\left(x\right)-NTA\sum\limits_{n=-\infty}^{\infty}\sum\limits_{l=\pm 1}\int \frac{d^2k}{4\pi^2}\ln\left[\left( \omega_n+i\omega_0\right)^2 +\mathcal{E}_{l}^2[\textbf A]\right]\right\}\right\} \nonumber\\
	&=&Z[\textbf{0}]\exp \left \{ -\beta T\sum_{\omega_n}\sum_{l=\pm 1}\int \frac{d^2k}{(2\pi)^2} \ln\left[\frac{ \left( \omega_n+i\omega_0\right)^2 +  \mathcal{E}_{l}^2[\textbf A]}{ \omega_n^2 +  \mathcal{E}_{l}^2[0]} \right ]\right \} ,
\label{8}
\end{eqnarray}
\end{widetext}
where $\omega_0= \frac{\mu_B \mu_0 H}{2\hbar}$ is the Zeeman coupling of the external field to the holes' spin and $\omega_n=(2n+1)\frac{\pi}{\beta}$, are the Matsubara  frequencies corresponding to the fermion integration.

We now perform the sums in the previous equation, using
\begin{widetext}
\begin{eqnarray}
  \sum_{n=-\infty}^\infty\frac{1}{ \left( \omega_n+i\omega_0\right)^2 +  \mathcal{E}_{l}^2}
= \frac{\beta}{4\mathcal{E}_{l}}
\left\{\tanh\left[\frac{\left[\mathcal{E}_{l}[\textbf A]+\hbar \omega_0\right ]}{2k_BT}\right] +\tanh\left[\frac{\left[\mathcal{E}_{l}[\textbf A]-\hbar \omega_0\right ]}{2k_BT}\right]\right \}
    \label{soma}
\end{eqnarray}
\end{widetext}

Inserting
 (\ref{soma}) in (\ref{8}), and using the rules of functional differentiation \cite{ecm2}, we obtain the average current: $\langle j^i \rangle$.
\begin{widetext}
\begin{eqnarray}
\langle j^i \rangle \left(\textbf{k}=0, \omega=0\right)=\frac{N}{2} \sum_{l =\pm 1} \frac{\partial   \mathcal{E}_{l}[\textbf A]}{\partial 
\textbf {A}^i} \left\{\tanh\left[\frac{\left[\mathcal{E}_{l}[\textbf A]+\hbar \omega_0\right ]}{2k_BT}\right] +\tanh\left[\frac{\left[\mathcal{E}_{l}[\textbf A]-\hbar \omega_0\right ]}{2k_BT}\right]\right \}.
\label{10}
\end{eqnarray}
\end{widetext}

To calculate the conductivity matrix, $\sigma^{ij}$, using (\ref{10z}) we need the two-point current correlator, hence we must take the derivative of $\langle j^i \rangle$ with respect to $\textbf {A}^j$, at $\textbf {k}=0$. 

We shall be primarily interested in the diagonal elements of the conductivity and resistivity matrices. For these 
\begin{widetext}
\begin{eqnarray}
\hspace{-0.5cm}\langle j^i j^i \rangle\left(\textbf{k}=0, \omega=0\right) = 
%\nonumber \\
\frac{N e^2 v^2 }{\mathcal{M}} \sum_{l,s=\pm 1}   
 \frac{\mathcal{M}+l\mu}{\sqrt{\Delta^2+\Big ( \mathcal{M} + l\mu \Big )^2}}\tanh\left (\frac{  \sqrt{\Delta^2+\Big ( \mathcal{M} + l\mu} \Big )^2+s \hbar\omega_0}{2k_BT}\right) .
\nonumber \\
\label{11}
\end{eqnarray}

%\begin{eqnarray}
%\langle j^i j^i\rangle\left(\textbf{k}=0, \omega=0\right) %=
 %N \sum_{l =\pm 1} \frac{\partial^2   \mathcal{E}_{l}[\textbf A]}{\partial \textbf {A}^i \partial \textbf {A}^i}\frac{\sinh\Big (\frac{  \mathcal{E}_{l}[\textbf A]}{k_BT}\Big)}{\cosh\Big (\frac{  \mathcal{E}_{l}[\textbf A]}{k_BT} \Big )+ \cosh\Big (\frac{  \hbar\omega_0}{k_BT}\Big )}.
 %\nonumber\\
%\label{10y}
%\end{eqnarray}

\end{widetext}
where, we used (\ref{7aaa}),  to define
\begin{equation}
    \mathcal{M}^2 \equiv M^2 + e^2v^2 \textbf{A}^2.
    \label{m}
\end{equation}
Considering that 
  \begin{equation}
    \textbf{A}^2= \frac{1}{4} \langle r^2 \rangle \left (\mu_0 H \right )^2,
    \label{m2}
\end{equation}
where we have replaced the square of the position vector by its average value, related to the de Broglie wavelength: 
\begin{equation}
\langle r^2 \rangle\simeq \left( \frac{\hbar}{m v}\right)^2 =  \left( \frac{\hbar}{m_ev}\right)^2 \left( \frac{m_e}{m}\right)^2 \end{equation}
where $m$ is the effective quasiparticle mass and $m_e$ the electron mass, we can express (\ref{m}) as
 
 \begin{equation}
 \mathcal{M}^2 = M^2 + e^2v^2 \textbf{A}^2= M^2 + \left( \frac{e\hbar}{2m_e}\right)^2 \lambda_2^2(\mu_0 H)^2
\end{equation}
 \begin{widetext}
 which implies
 \begin{equation}
\frac{\mathcal{M}}{k_BT}\equiv \frac{\sqrt{M^2+(\lambda_2\mu_B\mu_0H)^2}}{k_BT} = \sqrt{\left( \frac{M}{k_BT}\right )^2+ 
 \lambda_2^2 \left( \frac{\mu_B\mu_0 H}{k_BT}\right )^2},
\end{equation}
where $\mu_B=\frac{e\hbar}{2m_e}$ is the Bohr magneton, $\frac{\mu_B}{k_B}=0.671 \ K/T$ and $ \lambda_2 \simeq   \frac{m_e}{m}$.

%In order to obtain the DC conductivity per CuO$_2$ plane, we just divide by $N$.
The corresponding DC resistivity per CuO$_2$ plane, then, will be given by (we drop from now on, the $ij$-superscript).
\begin{widetext}
\begin{eqnarray}
\ \rho =\left(\frac{\sigma_{\text{DC}}}{N}\right)^{-1}=
	 \frac{ \mathcal{M}}{\hbar\beta V^{-1} e^2 v^2 \sum_{l,s=\pm 1}   
 \frac{\mathcal{M}+l\mu}{\sqrt{\Delta^2+\Big ( \mathcal{M} + l\mu \Big )^2}}\tanh\left (\frac{  \sqrt{\Delta^2+\Big ( \mathcal{M} + l\mu} \Big )^2+s \hbar\omega_0}{2k_BT}\right) 
 }.
 \label{rrr}
	\end{eqnarray}
	\end{widetext}
where $V=da^2$ is the volume of the primitive unit cell, per CuO$_2$ plane, with $d$ being the distance between planes, $a$ the lattice parameter and $v$, the characteristic velocity of the holes (such that for LSCO $\left(\hbar v/a\right)\approx 2.86\times 10^{-2} eV$~\cite{marino2020superconducting}).

	%In the SC phase, we have $\Delta \neq 0$ and $M =0$, implying that the resistivity at zero magnetic field vanishes %as it should:
	%\begin{eqnarray}
	%\rho_{SC}^{ij} 
	% &=&\frac{\delta^{ij} M}{\hbar\beta V^{-1} e^2 v^2  \left \{ \frac{2|\mu|}{\sqrt{\Delta^2+\mu^2}}\tanh\left %[\frac{\sqrt{\Delta^2+ \mu^2}}{2k_BT}  \right ] \right \}} \stackrel{M\rightarrow 0}{\longrightarrow 0}
	%\end{eqnarray}
%	\\
%\bigskip
 In the SC phase, we have $\Delta \neq 0$, $M=0$ and we can see, from (\ref{rrr}), that in the absence of an applied magnetic field, we have $\rho \rightarrow 0$ as a consequece of the fact  that, in this case $\mathcal{M}\rightarrow 0$ in (\ref{rrr}). By the same token, we are able to understand why the resistivity is no longer zero when an external magnetic field is applied: in this case, because of the magnetic field in (\ref{m}), we have  $\mathcal{M}\neq 0$  and the resistivity in (\ref{rrr}) does not vanish.\\
 \bigskip
 \\
\bigskip

{\bf 2.2) The Scaling Function}\\
\bigskip
\begin{widetext}

Outside the superconducting phases, we have $\Delta=0$, which leads to the following expression for the resistivity
\begin{eqnarray}
 \rho = \frac{Vk_B }{\hbar e^2 v^2 } \frac{\mathcal{M} T}{ \left[ \tanh\Big (\frac{ \mathcal{M} + \mu +  \hbar\omega_0}{2k_BT}\Big )+
  \tanh\Big (\frac{ \mathcal{M} + \mu -  \hbar\omega_0}{2k_BT}\Big )+ \tanh\Big (\frac{ \mathcal{M} - \mu +  \hbar\omega_0}{2k_BT}\Big )+
  \tanh\Big (\frac{ \mathcal{M} - \mu -  \hbar\omega_0}{2k_BT}\Big )\right ]}
  	\end{eqnarray}
  	
  	Using the identity,

\begin{eqnarray}
 \frac{2\sinh (a)}{\cosh (a )+ \cosh(b)}
= \tanh\left(\frac{a+b}{2}\right) +\tanh\left(\frac{a-b}{2}\right)
    \label{soma1}
\end{eqnarray}

  	for the 1st.+ 4th. and 2nd + 3rd. terms above, this can be rewritten as
  	
  	\begin{eqnarray}
 \rho&=&\frac{Vk_B }{\hbar e^2 v^2 } \frac{\mathcal{M} T}{ \left[\frac{\sinh\left (\frac{ \mathcal{M}  }{k_BT}\right )}{\cosh\left (\frac{ \mathcal{M} }{k_BT}\right ) + \cosh\Big (\frac{ \mu+ \hbar\omega_0}{k_BT}\Big )}+
 \frac{\sinh\left (\frac{ \mathcal{M}  }{k_BT}\right )}{\cosh\left (\frac{ \mathcal{M}  }{k_BT}\right ) + \cosh\Big (\frac{ \mu - \hbar\omega_0}{k_BT}\Big )  }\right ]} 
 \label{ro1}
 \end{eqnarray}
% \\ \nonumber
%&& = \frac{Vk_B }{\hbar e^2 v^2 } \frac{\mathcal{M} T}{  \tanh\Big (\frac{ \mathcal{M} + \mu +  \hbar\omega_0}{k_BT}\Big )+
 % \tanh\Big (\frac{ \mathcal{M} + \mu -  \hbar\omega_0}{k_BT}\Big )+ \tanh\Big (\frac{ \mathcal{M} - \mu +  \hbar\omega_0}{k_BT}\Big )+
  %\tanh\Big (\frac{ \mathcal{M} - \mu -  \hbar\omega_0}{k_BT}\Big )}
  	or
\begin{eqnarray}
 \rho&=&\frac{Vk_B }{\hbar e^2 v^2 } 
  \frac{\mathcal{M} T}{2\sinh\left (\frac{ \mathcal{M}  }{k_BT}\right )} \left\{\frac{\left [ \cosh\left (\frac{ \mathcal{M} }{k_BT}\right )+\cosh\left (\frac{ \mathcal{\mu} }{k_BT}\right )\cosh\left (\frac{  \hbar\omega_0 }{k_BT}\right ) \right]^2 -\left[\sinh\left (\frac{ \mathcal{\mu} }{k_BT}\right )\sinh\left (\frac{  \hbar\omega_0 }{k_BT}\right ] \right)^2 }{\cosh\left (\frac{ \mathcal{M} }{k_BT}\right )+\cosh\left (\frac{ \mathcal{\mu} }{k_BT}\right )\cosh\left (\frac{  \hbar\omega_0 }{k_BT}\right )}  \right \}
 \label{rho}
  	\end{eqnarray}
\end{widetext}
We can express the resistiviy in the presence of an applied magnetic field in terms of a three-variable scaling function $G(K_1, K_2, K_3)$, where 
\begin{equation}
    K_1=\frac{M}{k_BT}\ \ ;\ \ K_2=\frac{\mu}{k_BT}
    \ \ ;\ \ K_3=\frac{\mu_B\mu_0 H}{k_BT};
\end{equation}
namely

\begin{equation}
		\rho(x,T)=BT^2 G\left(\frac{M}{k_B T}, \frac{\mu}{k_B T},\frac{\mu_B\mu_0 H}{k_B T}\right),
		\label{eq_rho}
	\end{equation}
where
\begin{widetext}
\begin{equation}
 	G\left(K_1,K_2,K_3\right)=\frac{\sqrt{K_1^2+(\lambda_2 K_3)^2}}{2\sinh \left(\sqrt{K_1^2+(\lambda_2 K_3)^2}\right)}\ 
 	\left[ \frac{\left(\cosh \sqrt{K_1^2+(\lambda_2 K_3)^2} +\cosh K_2\cosh K_3\right)^2- \left(\sinh	K_2\sinh K_3\right )^2 }{\cosh \sqrt{K_1^2+(\lambda_2 K_3)^2} +\cosh K_2\cosh K_3} \right ]
 	\label{G}
 	\end{equation}
 		\end{widetext}
	and $B$ is given by
	\begin{equation}
		B=\frac{h}{e^2}\frac{d}{2\pi}\left(\frac{a}{\hbar v}\right)^2k_B^2.
		\label{eq_B}
	\end{equation}
For LSCO, we have $B_{LSCO}= 2.4457 \  n\Omega \text{cm}/K^2$ and, in general, we write $B=\lambda_1 B_{LSCO} $.

%\left( \frac{m_e}{m}\right)^2.
 \end{widetext}

%This can be rewritten as
%\begin{eqnarray}
%	 	 \rho&=&  \frac{Vk_B}{\hbar v^2e^2}T\frac{ %\mathcal{M}\left[\cosh\left(\frac{M}{k_BT}\right)+\cosh\left(\frac{\mu}{k_BT}\right)\right]}{2\sinh\left(\frac{M}{k_BT}\right)}.
%\end{eqnarray}
 Notice that in the zero field limit, $K_3\rightarrow 0$ and our expression for the resistivity reduces to the one in \cite{Res}.
 \ \ \ \ \  \\ \\
\bigskip
{\bf 2.3) The Strange Metal Phase}
\\
\bigskip

  Particularly interesting is the strange metal phase, where we have, both the SC and PG parameters vanishing: $\Delta=0$ and $M=0$. I
The chemical potential, conversely, scales with $T$, namely $\mu=D T$, where $D=2.69 \ eV/K$ \cite{Res}. 
Consequently, we will have
$K_1=0$, $K_2= D/k_B$, $K_3=\frac{\mu_B\mu_0 H}{k_BT}$. 
Combining these results in (\ref{ro1}), we 
can express the resistivity as
\begin{widetext}
\begin{eqnarray}
 \rho&=& \frac{\lambda_1 \lambda_2 BT^*TK_3  }{\sinh\left (\lambda_2 K_3\right )}\left[   \frac{1}{\cosh\left (\lambda_2K_3 \right )+
 \cosh\left( K_3 +D/k_B\right )} +
 %\\ \nonumber
 \frac{1}{\cosh\left (\lambda_2K_3 \right )+
 \cosh\left (K_3 -D/k_B\right )}\right ] .
\label{r}
	\end{eqnarray}
	where $T^*$ is the PG temperature.
\end{widetext}

%	We can express the above equation in terms of the scaling function $G$ of the critical variables $K_1=M_0/k_B T$, $K_2=\mu/k_B T$, 
%	and $K_3=\mu_0 H/k_B T$.
%	
%	given by
%	\begin{equation}
%		G\left(K_1,K_2\right)=K_1\frac{\cosh K_1 +\cosh K_2}{2\sinh \left(K_1\right)},
%		\label{eq_G}
%	\end{equation}
%	such that the resistivity becomes (From now on we will drop the $ij$ index)
%	\begin{equation}
%		\rho(x,T)=BT^2 G\left(\frac{M}{k_B T}, \frac{\mu}{k_B T}\right),
%		\label{eq_rho}
%	\end{equation}
%where the (almost universal) constant $B$ is
%	\begin{equation}
%		B=\frac{h}{e^2}\frac{d}{2\pi}\left(\frac{a}{\hbar v}\right)^2k_B^2\approx 0.37\times d  \  n\Omega \text{cm}/K^2,
%		\label{eq_B}
%	\end{equation}
%	where  $h/e^2\approx 25 812.807 \Omega$ is the resistance quantum and
 %$d$ is given in \AA\ -units.

%	This general form of the resistivity, whose dependence on the temperature ($T$) and on the doping parameter ($x$) has been made explicit, holds in all phases of the phase diagram of cuprates, except the SC one. The peculiar form of the resistivity in each of the different phases will be determined by the form the function $G\left(K_1,K_2\right)$ assumes in each phase.
	
%	\cite{phillips2012advanced, kirkpatrick15, phillips05, sachdev97}
\bigskip
{\bf 2.4) The Zero Magnetic Field Regime}
\\
\bigskip

In the $H\rightarrow 0$ limit, we have $K_3\rightarrow 0$. In this case, (\ref{G}) reduces to
\begin{eqnarray}
   &&  G\left(K_1,K_2,K_3\right)\longrightarrow 
     \\ \nonumber
 &&	G\left(K_1,K_2\right)=\frac{K_1}{2\sinh K_1}\left[\cosh K_1 + \cosh K_2 \right ] 
 	\label{G0}
 	\end{eqnarray}
 In the SM phase, where we also have $K_1=0$, accordingly, the scaling function becomes
\begin{eqnarray}
     G\left(K_1,K_2\right) =C\frac{T^*}{T}
   	\label{G1}
 	\end{eqnarray}
 	and the resistivity, according to (\ref{eq_rho}), becomes
 	\begin{eqnarray}
     \rho(T) = C T^* T
   	\label{G2}
 	\end{eqnarray}
\ \ \  \\ \\
\bigskip
{\bf 3) The Resistivity of LSCO }\\
\bigskip

Let us consider here a sample of LSCO, with a doping parameter $x=0.19$, which has a $T_c = 38.5 K$, that has been studied  in \cite{gg}. 

\begin{figure}
	[h]
	\centerline
	{
		%\figurename{TN}
		\includegraphics[scale=0.4]{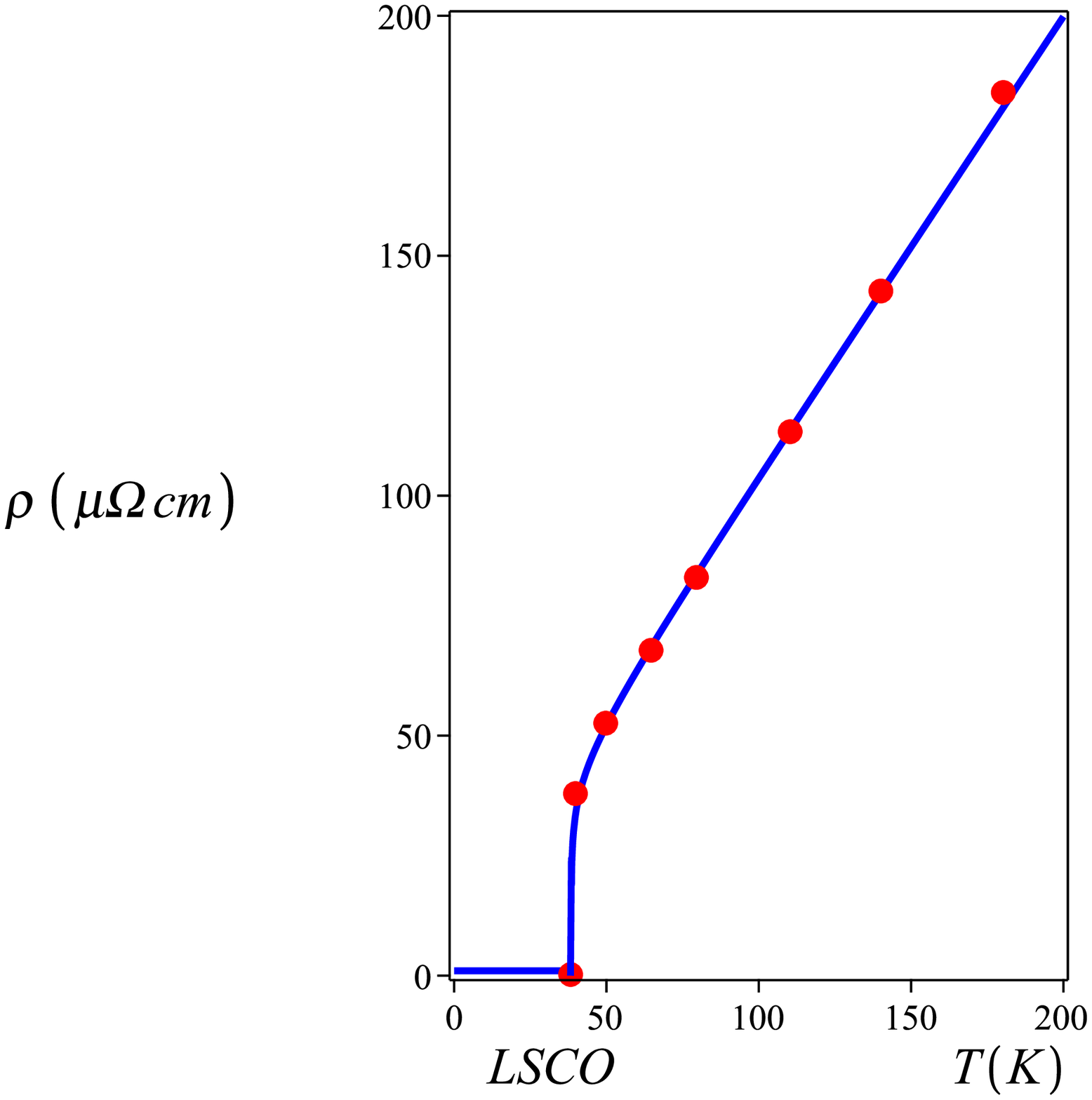}}
		%clip, 
		%angle=90, 
		%width=1.\textwidth]
		%\include{TN}
	
	\caption{Resistivity of LSCO, $\mu_0H=0$. The solid line is the plot of the theoretical expression derived from our theory, Eq. (\ref{r}) for a sample with $T_c=38.5K$ at zero magnetic field. Experimental data from \cite{gg}}
	\label{f1}
\end{figure}

\begin{figure}
	[h]
	\centerline
	{
		%\figurename{TN}
		\includegraphics[scale=0.4]{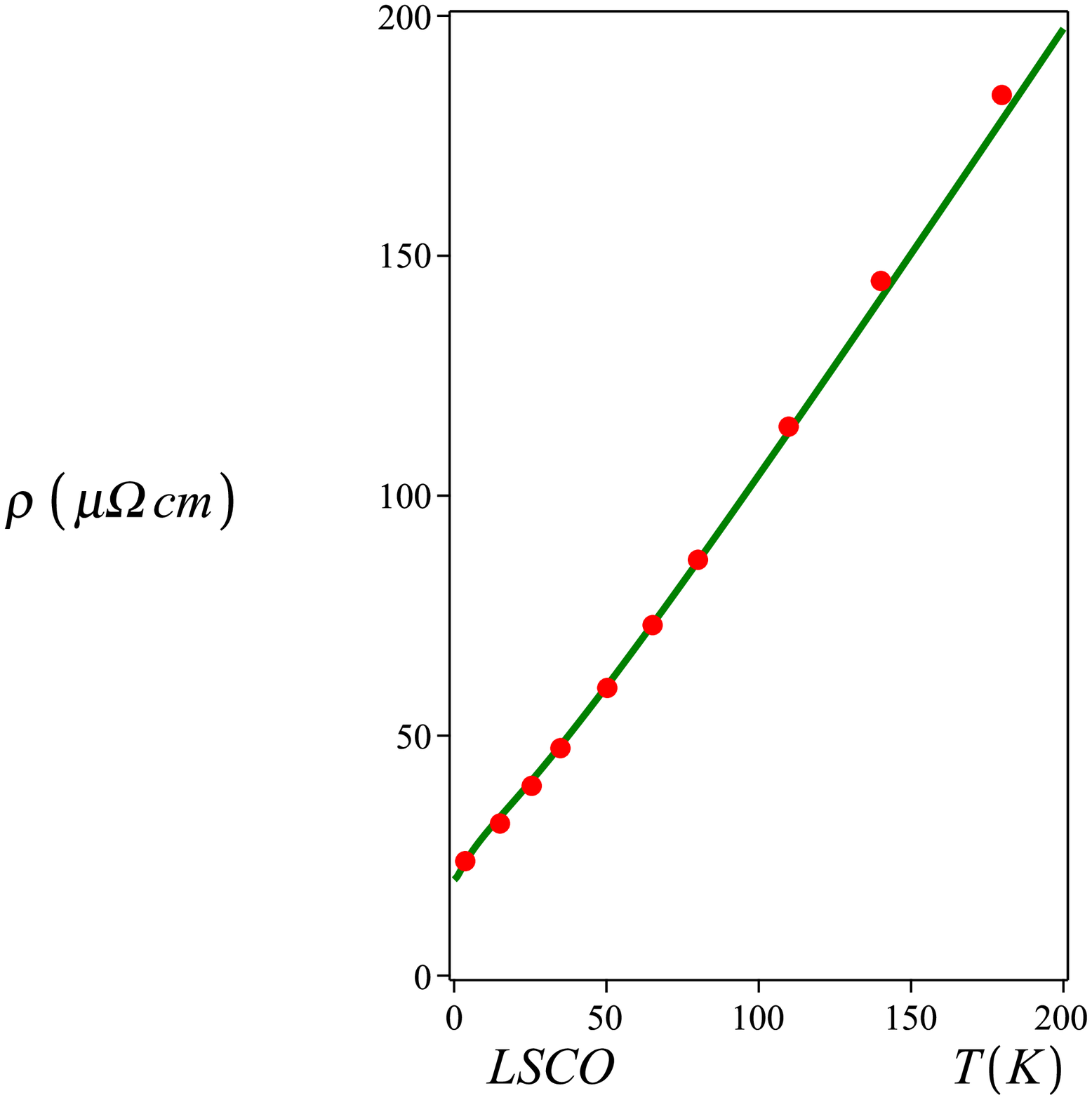}}
		%clip, 
		%angle=90, 
		%width=1.\textwidth]
		%\include{TN}
		\caption{Resistivity of LSCO. The solid line is the plot of the theoretical expression derived from our theory, Eq. (\ref{r}) for a sample with $T_c=38.5K$ at an applied  magnetic field of $50 T$. Experimental data from \cite{gg}}
	\label{f2}
\end{figure}

\begin{figure}
	[h]
	\centerline
	{
		%\figurename{TN}
		\includegraphics[scale=0.4]{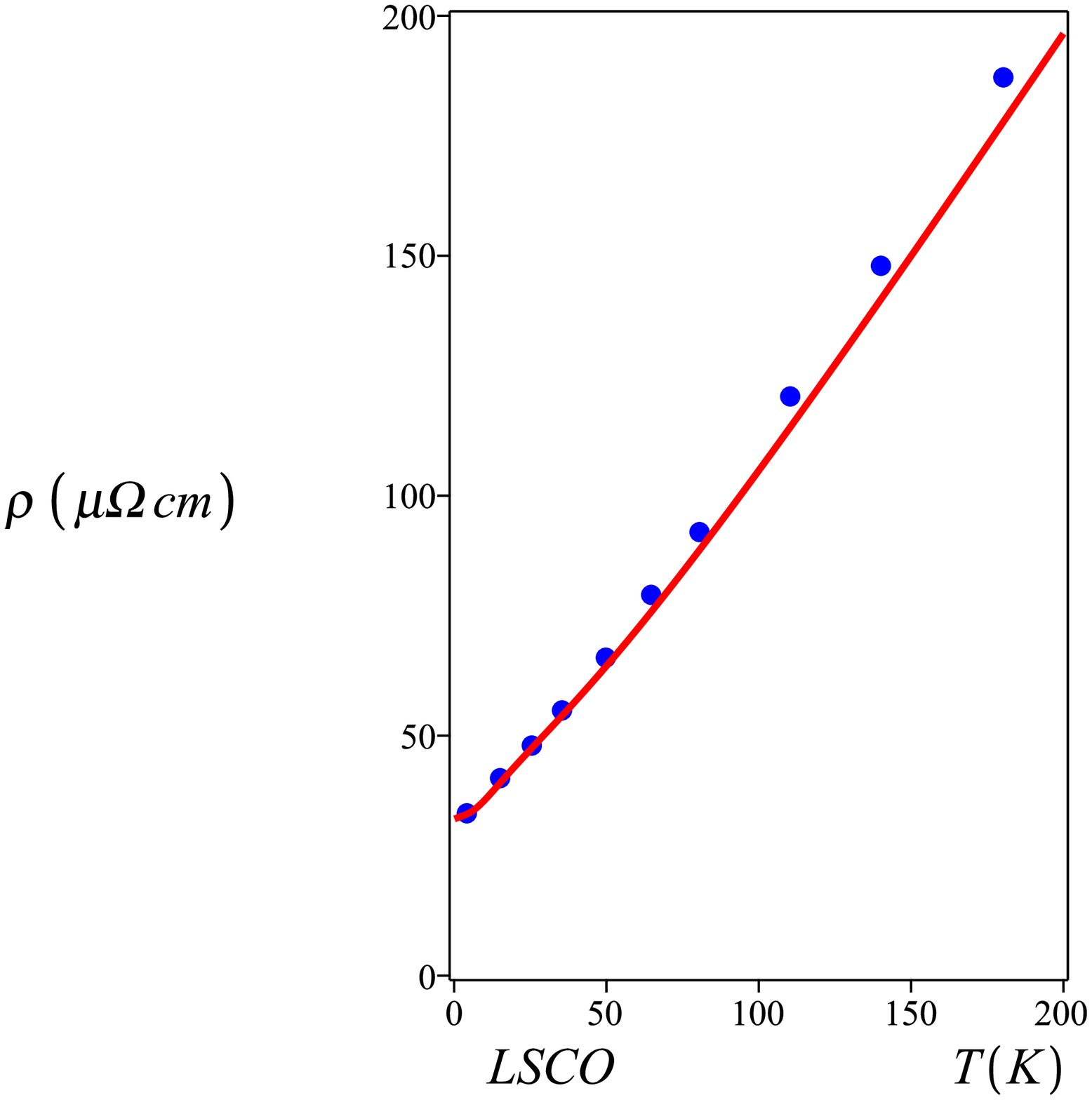}}
		%clip, 
		%angle=90, 
		%width=1.\textwidth]
		%\include{TN}
	
	\caption{Resistivity of LSCO. The solid line is the plot of the theoretical expression derived from our theory, Eq. (\ref{r}) for a sample with $T_c=38.5K$ at an applied  magnetic field of $80 T$. Experimental data from \cite{gg}}
	\label{f3}
\end{figure}

\begin{figure}
	[h]
	\centerline
	{
		%\figurename{TN}
		\includegraphics[scale=0.4]{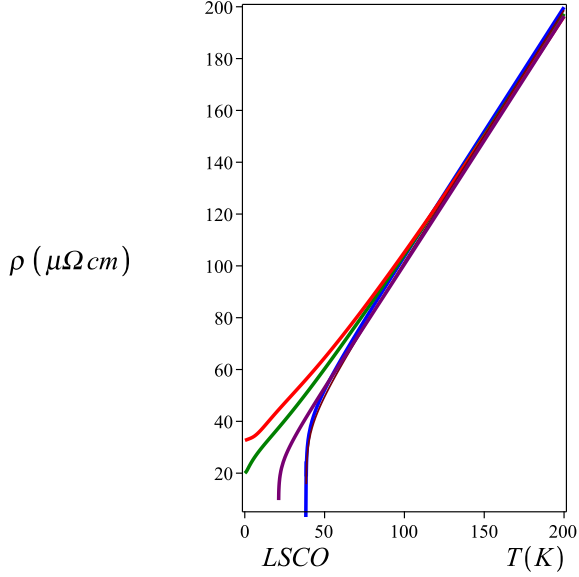}}
		%clip, 
		%angle=90, 
		%width=1.\textwidth]
		%\include{TN}
	
	\caption{Influence of an applied magnetic field on the resistivity of LSCO, for a sample with $T_c=38.5K$ at zero magnetic field  (blue line), $\mu_0H= 30T$ (purple line), $\mu_0H= 50T$ (green line) and $\mu_0H= 80T$ (red line). Observe that an applied magnetic field of $50T$ or more, completely destroys the SC phase.} \label{f4}
		\end{figure}

In Fig. \ref{f1} we plot our expression (\ref{r}), for the zero field resistivity (solid blue line), together with the experimental data from \cite{gg}. In Figs. \ref{f2} , \ref{f3}, we represent the curves corresponding to our expression (\ref{r}), respectively for an applied magnetic field of $50 T$ and $80 T$, along with the experimental data from \cite{gg}.
In Fig \ref{f4}, we depict the three curves together, along with the one for $30T$. Notice that magnetic fields of $50T$ and up are strong enough to destroy the SC phase.

We see that our expression for $\rho(T,H)$ is in excellent agreement with the experimetal data for LSCO.


\ \ \ \\
\\
\bigskip
{\bf 4) The Magnetoresistivity of Bi2201}\\
\bigskip

Let us now consider our general expression for the resisivity, in the SM phase, Eq. (\ref{r}), taken as a function of the applied magnetic field $H$. Let  us apply it for the sample of Bi2201, having $T_c \simeq 1K$ at  a fixed temperature $T=4.2K$ studied in \cite{Hussey2020}.

According to our expression for the SC transition temperature of cuprates \cite{marino2020superconducting,Res}, for Bi2201 a critical SC temperature of $T_c\simeq 1K$ corresponds to a stoichiometric doping parameter $x= 0.377$. 

Then, according to our expression for the PG temperature $T^*$ \cite{marino2020superconducting,Res} of cuprates, such doping parameter corresponds to $T^*=3.15K$. The sample of Bi2201, studied in \cite{HusseyTl} at a temperature of $T=4.2K$, therefore must be in the Strange Metal phase, where $M=0$ and $\mu=DT$ \cite{marino2020superconducting,Res}.

\begin{figure}
	[h]
	\centerline
	{
		%\figurename{TN}
		\includegraphics[scale=0.4]{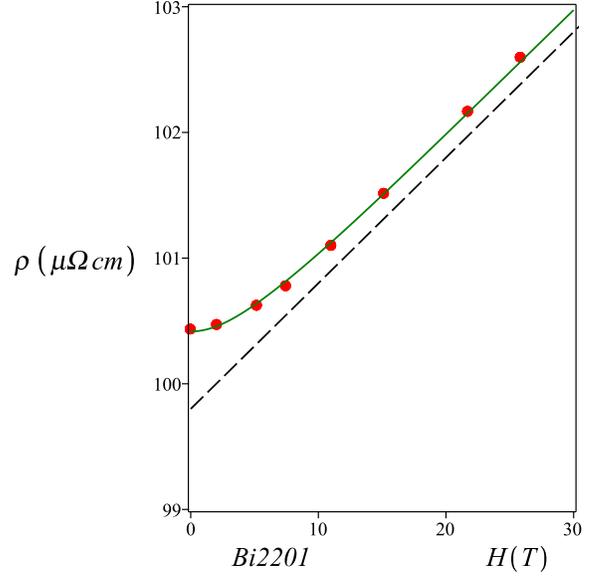}
		%clip, 
		%angle=90, 
		%width=1.\textwidth]
		%\include{TN}
	}
	\caption{Magnetoresistance of Bi2201. Our theoretical expression, derived from first principles (green line), accurately describes the experimental result obtained in \cite{Hussey2020}, for a sample of Bi2201 with $T_c\simeq 1K$, which corresponds to a pseudogap temperature  $T^*=3.15 K$ \cite{marino2020superconducting}. The measurement is made at a temperature $T=4.2K$, which is larger than $T^*$, implying that the material is in the SM phase. The (linear) dashed line is added just to emphasize the crossover of the dependency of $\rho$ with $H$.}
	\label{f5}
\end{figure}

Using our expression (\ref{r}) at a fixed temperature of $T=4.2K$ and choosing $\lambda_1= 25.32$, $\lambda_2=3$ and a residual resistivity $\rho_0=100 \mu\Omega cm$, we obtain the curve depicted in green in Fig. \ref{f5}. The experimental data are from \cite{Hussey2020}

\begin{figure}
	[h]
	\centerline
	{
		%\figurename{TN}
		\includegraphics[scale=0.4]{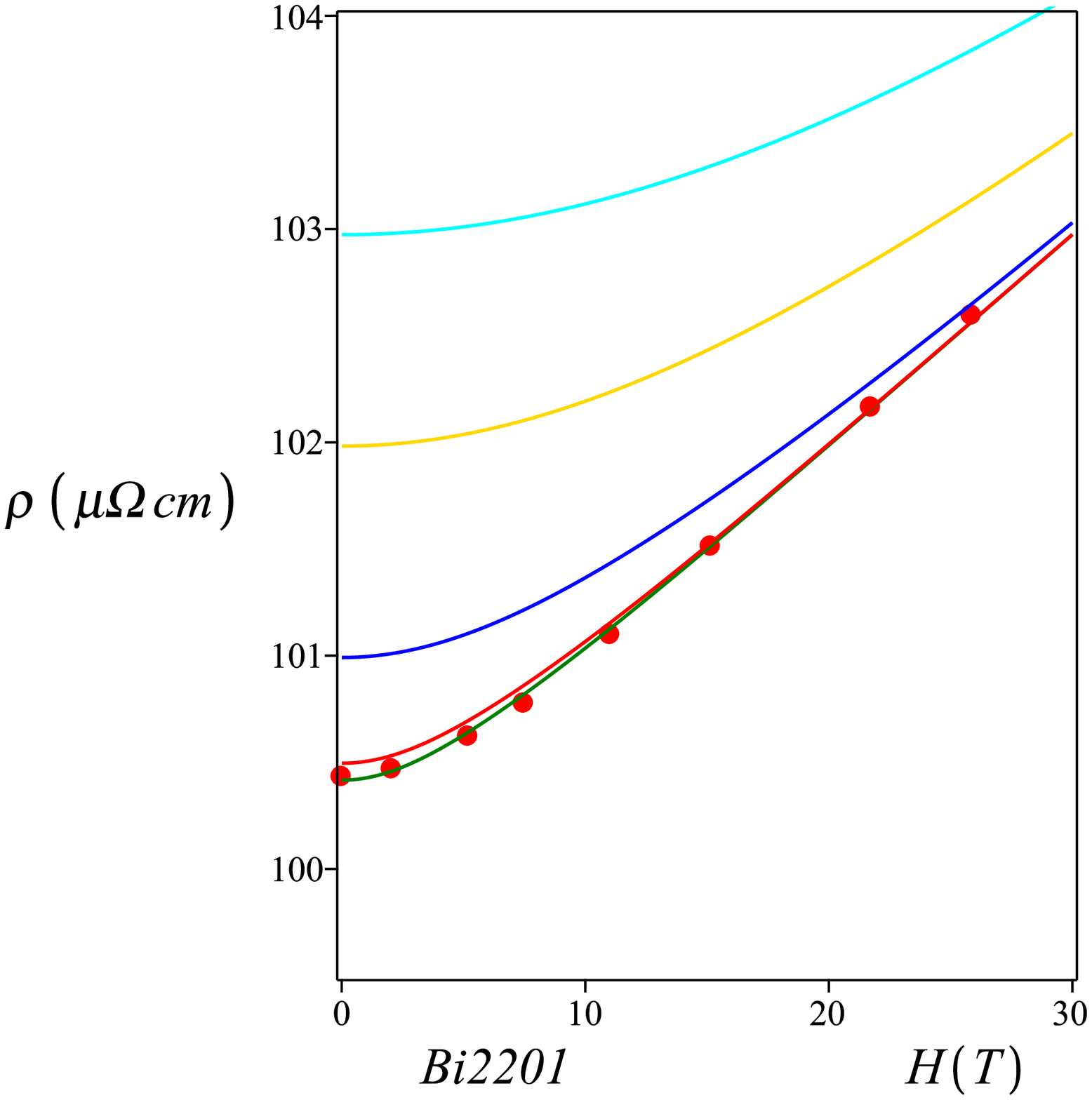}
		%clip, 
		%angle=90, 
		%width=1.\textwidth]
		%\include{TN}
	}
	\caption{Magnetoresistance of Bi2201, for the same sample of Fig. \ref{f5}. The different curves represent our analytical result corresponding to temperatures of $4.2K$ (green), $5K$ (red), $10K$ (blue), $20K$ (gold) and $30K$ (cyan). Experimental data are for the $4.2K$ sample (same as in Fig. \ref{f5}.}
	\label{f6}
\end{figure}
In Fig. \ref{f6} we show the magnetoresistance curves for the same sample of B12201 at different temperatures.

\bigskip
{\bf 5) The Resistivity of Tl2201 and the Location of the QCP}
\\
\bigskip

Let us take the case of Tl2201, in order to address the issue of the power-law dependence of the zero field resistivity near the SC dome in OD cuprates. As it turns out, knowledge of such power-law will enable to clarify the the issue concerning the location of the QCP associated to the SM phase. For this purpose, we are going to use the results obtained in \cite{Res}, according to which, we have the following power-law regimes for the resistivity just outside the SC dome: Strange Metal (SM), Fermi Liquid (FL), Crossover (C) :
\begin{eqnarray}
&& \mathrm{SM} \ -\ \rho\propto T
\\ \nonumber
&& \mathrm{FL} \ -\ \rho\propto T^2
\\ \nonumber
&& \mathrm{C} \ -\ \rho\propto T^{1+\delta}\ \ \ \delta\in [0,1]
\\ \nonumber
\label{ff}
\end{eqnarray}
We also recall that the resistivity behavior in the upper PG phase shares the $T$-linear behavior with the SM phase \cite{Res}.

The scaling function $G\left(K_1,K_2\right)$ has the following types of behavior in each of the regions above \cite{Res} 
\begin{eqnarray}
&& \mathrm{SM} \ - \  G\propto \frac{T^*}{T}
\\ \nonumber
&& \mathrm{FL} \ -\ G\propto C
\\ \nonumber
&& \mathrm{C} \ -\ G\propto \left(\frac{T^*}{T}\right)^{1-\delta}\ \ \ \delta\in [0,1]
\\ \nonumber
\label{ss}
\end{eqnarray}

Let us consider now, the two following scenarios for the phase diagram of cuprates, which we illustrate for the case of Tl2201.

\begin{figure} 
	[h]
	\centerline
	{
		%\figurename{TN}
		\includegraphics[scale=0.4]{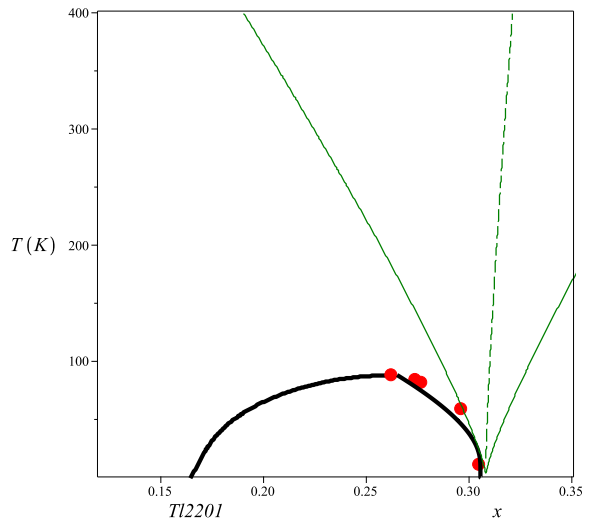}
		%clip, 
		%angle=90, 
		%width=1.\textwidth]
		%\include{TN}
	}
	\caption{Phase diagram of Tl2201. Scenario I where the QCP is located at the edge of the SC dome, in the OD region. The green lines delimit the SM region,on the left of the dashed line and the Crossover region, on the right. The FL region is found below the second green line. Experimental data from \cite{honma}.}
	\label{f12}
\end{figure}

\begin{figure}
	[h]
	\centerline
	{
		%\figurename{TN}
		\includegraphics[scale=0.4]{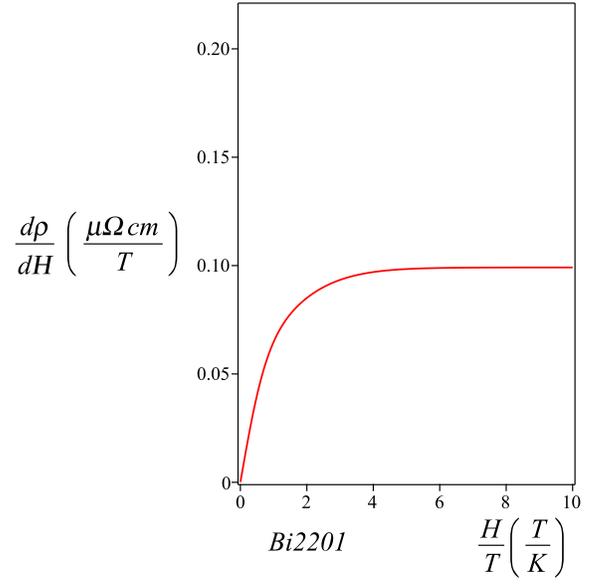}
		%clip, 
		%angle=90, 
		%width=1.\textwidth]
		%\include{TN}
	}
	\caption{Magnetoresistance derivative for Bi2201, plotted as a function of $H/T$.}
	\label{f11}
\end{figure}

In the first scenario (I), depicted in Fig. \ref{f11}, the quantum critical point (QCP) where the pseudogap line $T^*(x)$ ends, is located precisely at the edge of the SC dome, while in the second scenario, (II) which is depicted in Fig. \ref{f12}, the quantum critical point (QCP) is located inside the SC dome.

\begin{figure}
	[h]
	\centerline
	{
		%\figurename{TN}
		\includegraphics[scale=0.4]{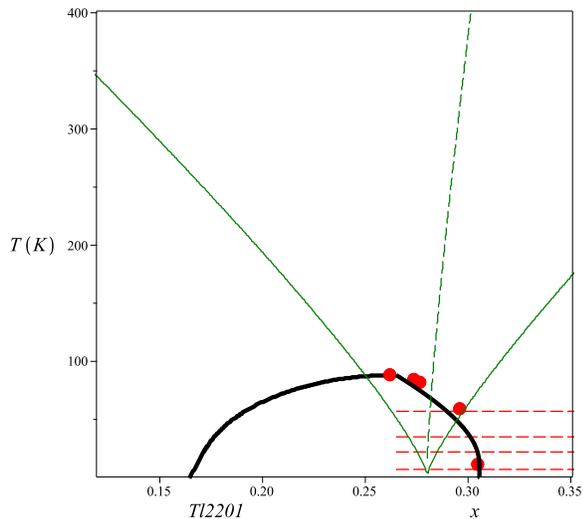}
		%clip, 
		%angle=90, 
		%width=1.\textwidth]
		%\include{TN}
	}
	\caption{Phase diagram of Tl2201. Scenario II, where the QCP is inside of the SC dome.The green lines delimit the SM region, on the left of the dashed line and the Crossover region, on the right. The FL region is found below the second green line. In any case, inside the SC dome, the SC state should be energetically favorable. The four dashed red lines mark the $T_c$ of the Tl2201 sample, respectively, 57K,35K,22K and 7K. Experimental data from \cite{honma}.}
	\label{f13}
\end{figure}

 Attentive inspection of these phase diagrams allows for the following conclusion. In the first scenario, the transition from the SC dome, in the OD region, always leads to a linear $\rho\propto T$ behavior of the resistivity. In the second scenario, conversely, according to Fig. \ref{f13}, the resistivity behavior depends on where we cross the SC dome: if we do it below the green line on the right-hand-side, we shall have a $\rho\propto T^2$ behavior. When we cross the SC dome between the dashed line and the green line on the right-hand-side, we shall have, conversely, a $\rho\propto T^{1+\delta}$, super-linear behavior. Finally, when we cross the SC dome in between the green line on the left-hand-side and the dashed line, we shall have a linear behavior, $\rho\propto T$. In any of the three cases, however, as we raise the temperature, we will eventually reach a $T$-linear behavior of the resistivity.
 
 From the behavior of the resistivity of a given cuprate material in the OD region one may infer about what type of scenario we will observe in its phase diagram, concerning especially the PG temperature line $T^*(x)$ and the position of the QCP.
 
 For the case of LSCO, for instance, the behavior exhibited in Fig. \ref{f1}, strongly suggests that scenario I applies to this material.
 
 Let us consider now the case of Tl2201. We evaluated the zero field resistivity,just above the SC transition for four samples, having, respectively, transition temperatures $T_c=7K,22K,35K,57K$. We did the calculation using (\ref{eq_rho}), with the different scaling functions in (\ref{ss}). The blue curves were obtained by using the scaling function of the FL phase. The red curves, conversely, were obtained with the scaling function of the Crossover.
 
\begin{figure}
	[h]
	\centerline
	{
		%\figurename{TN}
		\includegraphics[scale=0.4]{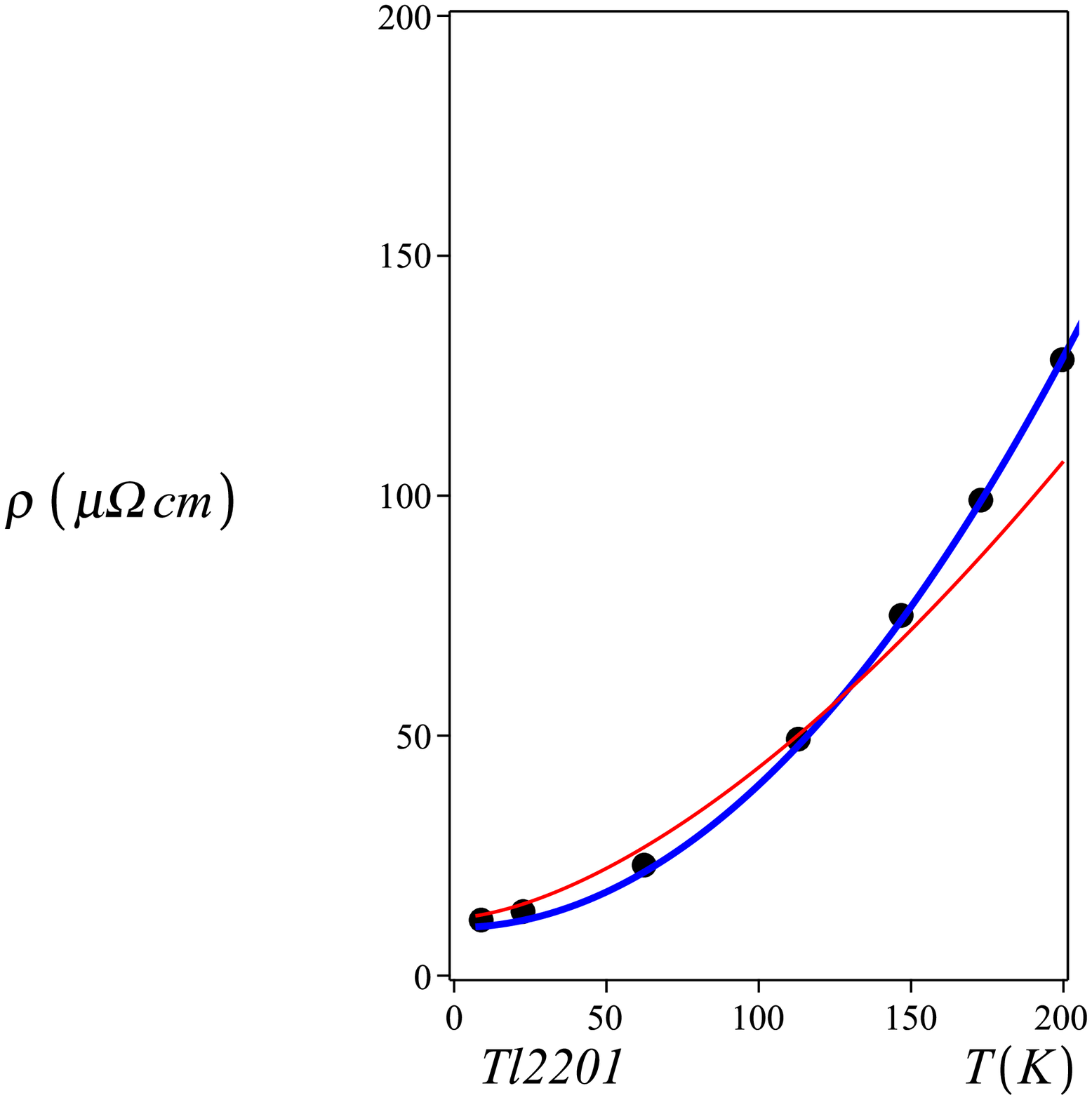}
		%clip, 
		%angle=90, 
		%width=1.\textwidth]
		%\include{TN}
	}
	\caption{Resistivity of Tl2201 at zero magnetic field, for a sample with $T_c=7K$. The blue line is our theoretical expression, calculated with the scaling function appropriate for the FL phase, while the red line would be the result, should we did the calculation with the one corresponding to the Crossover. Experimental data from \cite{HusseyTl,H1,H2,H3}.}
	\label{f7}
\end{figure}

\begin{figure}
	[h]
	\centerline
	{
		%\figurename{TN}
		\includegraphics[scale=0.4]{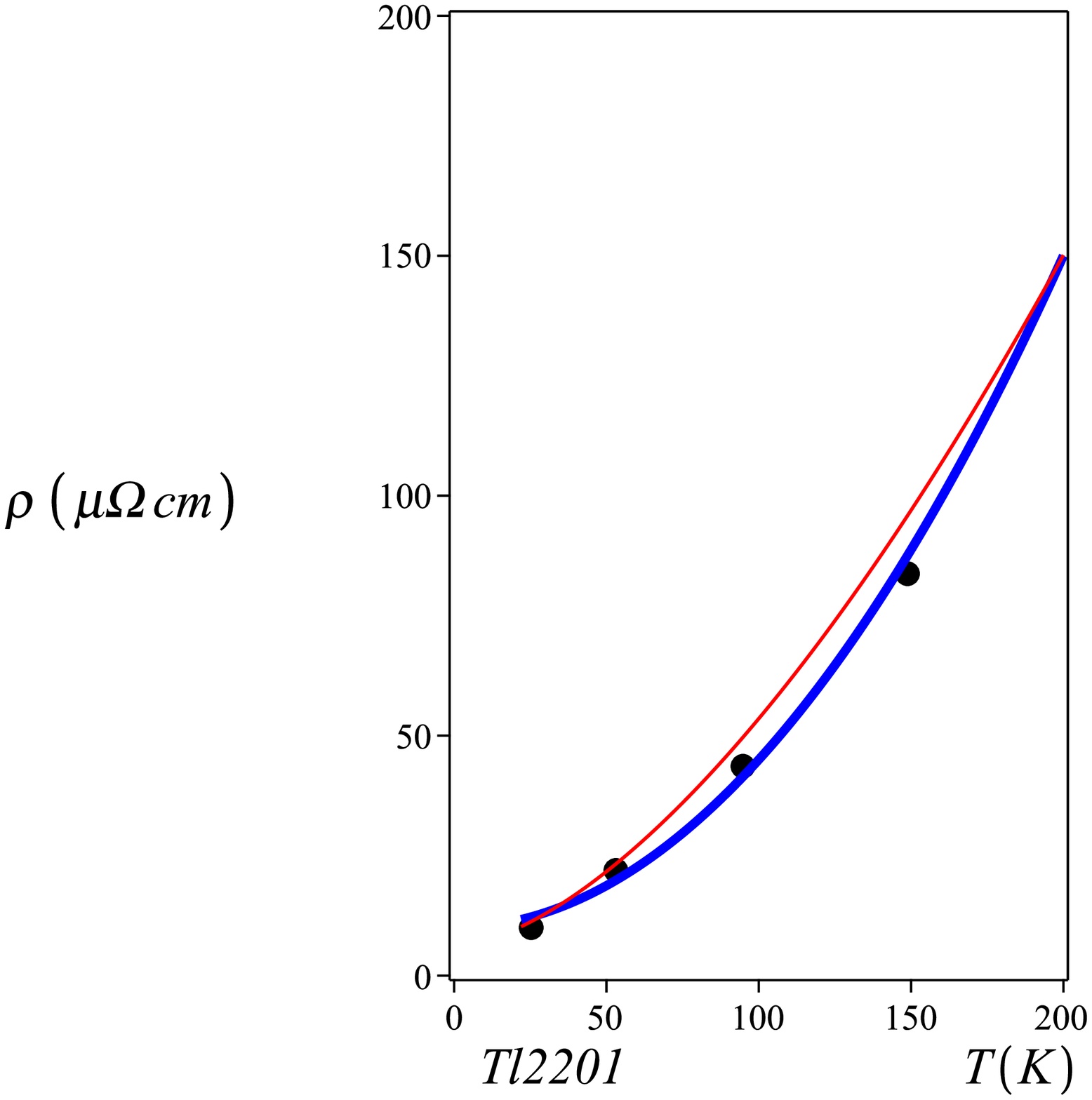}
		%clip, 
		%angle=90, 
		%width=1.\textwidth]
		%\include{TN}
	}
	\caption{Resistivity of Tl2201 at zero magnetic field, for a sample with $T_c=22K$. The blue line is our theoretical expression, calculated with the scaling function appropriate for the FL phase, while the red line would be the result, should we did the calculation with the one corresponding to the Crossover. Experimental data from \cite{HusseyTl,H1,H2,H3}.}
	\label{f8}
\end{figure}

\begin{figure}
	[h]
	\centerline
	{
		%\figurename{TN}
		\includegraphics[scale=0.4]{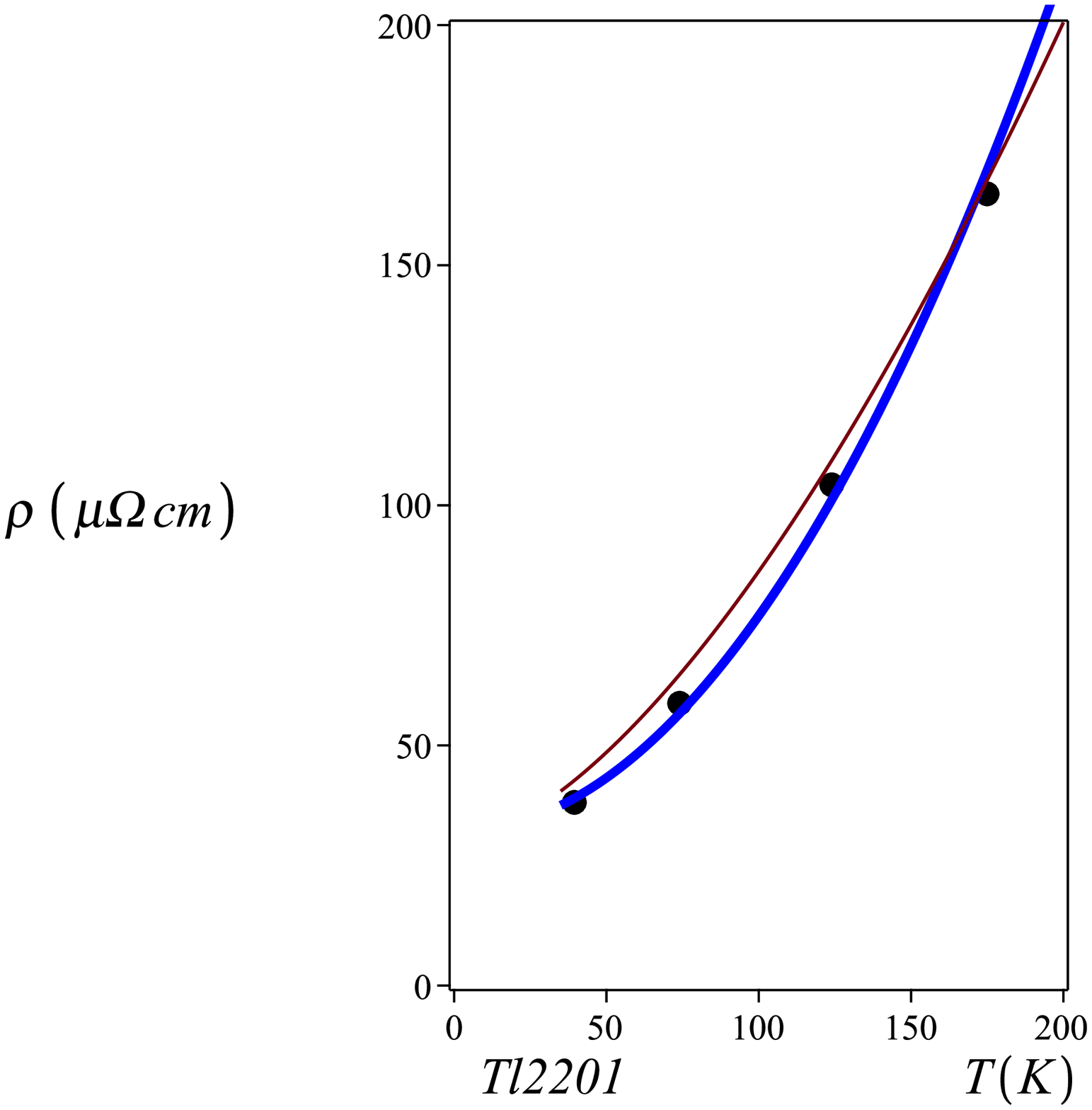}
		%clip, 
		%angle=90, 
		%width=1.\textwidth]
		%\include{TN}
	}
	\caption{Resistivity of Tl2201 at zero magnetic field, for a sample with $T_c=35K$. The blue line is our theoretical expression, calculated with the scaling function appropriate for the FL phase, while the red line would be the result, should we did the calculation with the one corresponding to the Crossover. Experimental data from \cite{HusseyTl,H1,H2,H3}.}
	\label{f9}
\end{figure}

\begin{figure}
	[h]
	\centerline
	{
		%\figurename{TN}
		\includegraphics[scale=0.4]{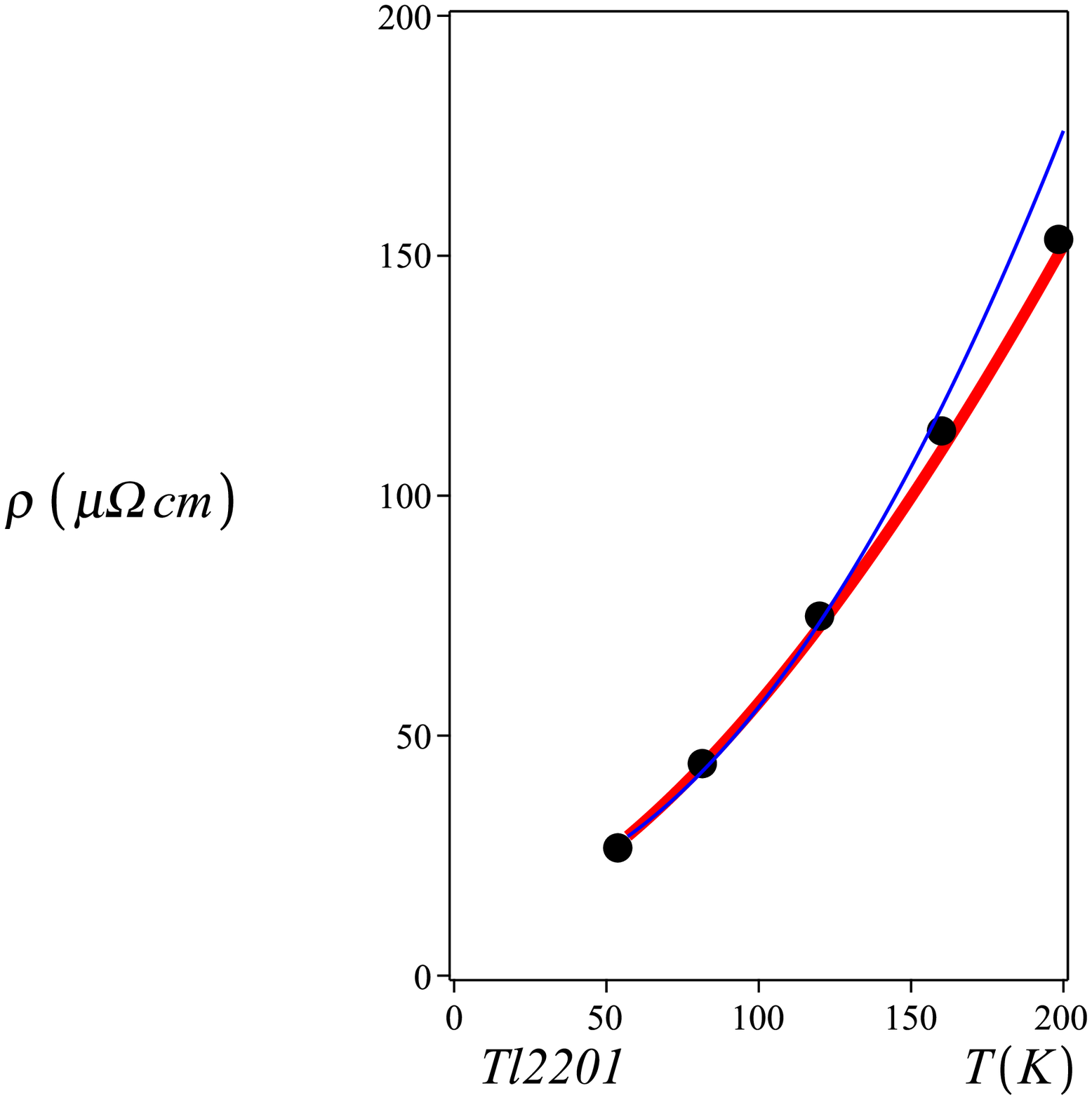}
		%clip, 
		%angle=90, 
		%width=1.\textwidth]
		%\include{TN}
	}
	\caption{Resistivity of Tl2201 at zero magnetic field, for a sample with $T_c= 57K$. The red line is our theoretical expression, calculated with the scaling function appropriate for the Crossover, while the blue line would be the result, should we did the calculation with the one corresponding to the FL phase. Experimental data from \cite{HusseyTl,H1,H2,H3}.}
	\label{f10} 
	\end{figure}

 The result, compared with experimental data of \cite{HusseyTl}, is shown in Figs. \ref{f7}, \ref{f8}, \ref{f9}, \ref{f10}. It shows unequivocally that the blue curves are the ones that correctly describe the resistivity of TL2201, for the 7K,22K and 35K samples of Tl2201 while the red curve correctly describe the resistivity of the 57K sample. We conclude that this sample undergoes the SC transition into the Crossover region while the other three samples do it from SC to FL phases. Remarkably, we can confirm the previous conclusions by visual inspection of the phase diagram in Fig. \ref{f13}, where the four red dashed lines represent the above samples of Tl2201.
   
   The results above, consequently, strongly suggest that scenario II applies for Tl2201.
 
%\begin{widetext}


%	\end{widetext}

\bigskip

{\bf 6) Quadrature and Scaling}\\
\bigskip

It was pointed out in \cite{Hussey2020} the existence of a crossover in the magnetoresistance in cuprates, from a quadratic behavior at low fields to a linear behavior in the high-fields regime. Our theoretical expression reproduces the experimentally observed crossover (see Figs. \ref{f5}, \ref{f6}).

This type of behavior, in  materials such as electron doped cuprates and iron pnictides is usually ascribed to a quadrature scaling of the MR supposed to be associated to quantum critical phases. 

In the case of cuprates, however, the same study of the magnetoresistivity in the SM phase indicates that the quadrature scaling is violated, in spite of the $H$-field crossover.
Moreover, the MR field derivative is shown to scale as function of $H/T$.

Our results indicate that despite exhibiting the quadratic to linear crossover, which is observed experimentally, the resistivity of cuprates in the Strange Metal phase does not show a quadrature scaling dependence. Rather it depends on $H$ and $T$, through the function in (\ref{r}), which was derived from our general theory for the cuprates \cite{marino2020superconducting,Res}.

In Fig. \ref{f11}, we display the field derivative of our expression (\ref{r}), plotted as a function of the ratio $H/T$, namely, for $y=\frac{H}{T}$,
\begin{equation}
\frac{\partial\rho_{SM}(H,T)}{\partial H}=\frac{\partial\rho_{SM}(y,T)}{\partial y}\frac{\partial y}{\partial H}=\frac{T}{T}\frac{df(y)}{dy}=f'(y),
\label{rr}
\end{equation}
where we used that $\rho_{SM}(H,T)=T f(y)$.

The resulting expression has precisely the form of the collapsed experimental data exhibited in \cite{Hussey2020}, which indicates the correction of the magnetoresistance derived from our theory.

\bigskip

{\bf 7) Conclusion}\\
\bigskip

We have derived, from our recently proposed theory for the high-Tc cuprates, an analytic expression for the resistivity in the presence of an external magnetic field. This shows an excellent agreement with the experimental data for the resistivity of LSCO at different values of the applied magnetic field.

The associated MR presents the crossover from parabolic to linear dependence despite the fact that it does not satisfy a quadrature scaling. Yet, the magnetic field derivative of the magnetoresistance presents a $H/T$ scaling, in complete agreement with the results of the experiments performed in \cite{Hussey2020}. 

We introduced a method to determine whether the QCP associated to the pseudogap temperature $T^*$ and the SM phase is located inside or outside (at the edge) of the SC dome. This is based on the observation of the power-law behavior of the resistivity, as a function of $T$, just above $T_c$, in the OD region. Our results indicate that the QCP is inside the dome for Tl2201 and on its very edge, for LSCO.

\bigskip
\vfill\eject

\clearpage
{\bf Acknowledgments}

E. C. Marino was supported in part by CNPq and by FAPERJ. R. Arouca acknowledges funding from the Brazilian Coordination for the Improvement of Higher Education Personnel (CAPES) and from the Delta Institute for Theoretical Physics (DITP) consortium, a program of the Netherlands Organization for Scientific Research (NWO) that is funded by the Dutch Ministry of Education, Culture and Science. \\
\bigskip
\\
\bigskip

{\it Corresponding author: ECM (marino@if.ufrj.br)}

\bibliography{cuprates}% Produces the bibliography via BibTeX.

\end{document}